\documentclass[preprint2]{aastex}

\shorttitle{A {\it BeppoSAX} Observation of the IC1262 Galaxy Cluster}

\shortauthors{Hudson, Henriksen, and Colafrancesco}

\begin{document}

\title{A {\it BeppoSAX} Observation of the IC1262 Galaxy Cluster}

\author{Daniel S. Hudson}
\affil{Physics Department, University of Maryland,
Baltimore County, 	Baltimore, MD 21250}

\author{Mark J. Henriksen\altaffilmark{1}}
\affil{Joint Center for Astrophysics, Physics Department, University of Maryland,
Baltimore County, 	Baltimore, MD 21250}

\author{Sergio Colafrancesco}
\affil{Osservatorio Astronomico di Roma, Via Frascati, 33 I-00040,  Monteporzio (Roma)  
ITALY}

\altaffiltext{1}{Laboratory for High energy Astrophysics, NASA/GSFC}

\begin{abstract}

We present an analysis of {\it BeppoSAX}  observations of the IC1262 galaxy cluster and report the first temperature and abundance measurements, along with preliminary indications of diffuse, nonthermal emission.  By fitting a 6$\arcmin$ ($\sim$360 {\it h$_{50}^{-1}$} kpc) region with a single Mewe-Kaastra-Liedahl model with photoelectric absorption, we find a temperature of 2.1 - 2.3 keV, and abundance of 0.45 - 0.77 (both 90\% confidence).  We find the addition of a power-law component provides a statistically significant improvement ({\it F}-test = 90\%) to the fit.  The addition of a second thermal component also improves the fit but we argue that it is physically implausible.  The power-law component has a photon index ($\Gamma_{X}$) of 0.4 - 2.8 and a nonthermal flux of (4.1  - 56.7) $\times$ 10$^{-5}$ photons cm$^{-2}$ s$^{-1}$ over the 1.5 - 10.5 keV range in the Medium Energy Concentrator spectrometer  detector.  An unidentified X-ray source found in the {\it ROSAT} High Resolution Imager observation ($\sim$0$\arcmin$.9 from the center of the cluster) is a possible explanation for the nonthermal flux; however, additional evidence of diffuse, nonthermal emission comes from the NRAO VLA Sky Survey and the Westerbork Northern Sky Survey radio measurements, in which excess diffuse, radio flux is observed after point-source subtraction.  The radio excess can be fitted to a simple power law with a spectral index of $\sim$1.8, which is consistent with the nonthermal X-ray emission spectral index.  The steep spectrum is typical of diffuse emission and the size of the radio source implies that it is larger than the cD galaxy and not due to a discreet source.

\end{abstract}


\section{Introduction}
Since the discovery of nonthermal emission in galaxy clusters, in the form of a radio halo in Coma \citep{will}, there has been a debate about the nature of its origin.  \citet{jaf} demonstrated that although Active Galactic Nuclei (AGNs) in clusters could explain the numbers of relativistic electrons observed, this model could not explain their ability to diffuse through the entire cluster within their short ($\sim$10$^{8}$ yr) radiative lifetime.  \citet{hol} proposed that primary electrons are able to diffuse at velocities greater than the Alfv\'{e}n velocity and therefore could diffuse from point sources throughout the ICM within their radiative lifetime.  There has also been some discussion that secondary electrons could explain the diffuse relativistic electron  population.  Secondary electrons are produced in proton-proton interactions from a population of relativistic protons \citep{den}, which have a relatively long radiative lifetime ($\sim$10$^{10}$ yr) \citep{blasi01}.  The majority of authors, however, have looked to in-situ re-acceleration models from MHD turbulence or shocks \citep[for example]{eil} to explain diffuse nonthermal emission from clusters.  Mergers are the most popular theory for the origin of these shocks, since they provide the energy needed to produce a radio halo \citep{trib,burns01,burns02}, and are able to transport electrons via bulk flows \citep{roe}.

\citet{reph} correlated synchrotron emission to X-ray emission by demonstrating that cosmic microwave background (CMB) photons will inverse Compton scatter (ICS) off relativistic electrons, and emerge in the X-ray regime.  The relationship between the X-ray and radio flux is determined by the strength of the cluster magnetic field.  With only $\sim$30 clusters known to have radio halos \citep[and references therein]{giov02}, and three hot, rich clusters: Coma \citep{fusco01}, A2256 \citep{fusco02}, and A2199 \citep{kaa01}, and one group: HCG 62 \citep{fuka}, with detected nonthermal X-ray emission, it is difficult to constrain the current models.  Detection of nonthermal emission in hot clusters is complicated because the X-ray emission is dominated by thermal emission in the 0.5 to 25 keV energy band, which is why the Phoswich Detection System (PDS) has been required to make nonthermal X-ray detections in hot clusters.  In cool clusters, such as IC1262 ({ \it kT} $\sim$2keV), any nonthermal emission should be detectable in the 1.5 - 10.5 keV range and observable in the Middle Energy Concentrator Spectrometer (MECS), which unlike the PDS, has spatial capabilities.  Therefore, cool clusters provide an opportunity to add additional constraints, because the spatial information from the MECS can be used to constrain the regions of nonthermal emission, and reduce the chances of contamination by AGNs. 

The IC1262 galaxy cluster (also Zwicky Cluster 1728.5+4353) is a  poor cluster of galaxies at a red-shift of z = 0.0343 \citep{col}, so that an angular distance of 1$\arcmin$ corresponds to 60 {\it h$_{50}^{-1}$} kpc.  The cluster is so-named because it is dominated by the X-ray bright cD galaxy, IC1262 (z = .0328 \citep{weg}).  The cluster was observed by the HRI in March 1997 for $\sim$26 ksec, and {\it BeppoSAX} in February 1999 for $\sim$100 ksec.  \citet{trin} analyzed the HRI data, mainly focusing on the cD galaxy.  

Throughout this letter, we assume a Hubble Constant of H$_{0}$ = 50 {\it h$_{50}$} km s$^{-1}$ Mpc$^{-1}$ and q$_{0}$ = $\case{1}{2}$.  Quoted confidence intervals are at a 90\% level, unless otherwise specified. 

\section{Observations and Methods}
\begin{figure}[tbp]
\includegraphics[scale=0.4,angle=0]{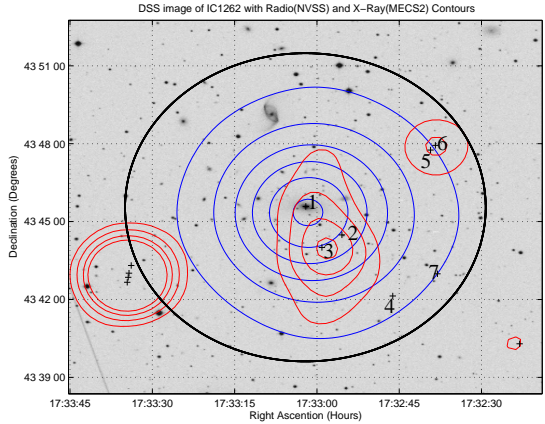}
\figcaption[f1.eps]
{A DSS image of IC1262 with MECS2 X-ray contours(blue), NVSS radio contours(red), and 6$\arcmin$ extraction region(black).  Contour levels are: 0.5,1.0,1.5,2.0,2.5,3.0 $\times$ 10$^{-5}$ photons cm$^{-2}$ s$^{-1}$ arcmin$^{-2}$. Radio contours are 1.35, 2.70, 4.50, 6.30 mJy beam$^{-1}$).  Numbered crosses represent the seven reported FIRST sources in the 6$\arcmin$ region.\label{fig1}}

\end{figure}

In order to produce a smoothed image from the MECS data, we first needed to create an exposure map.  The SAXDAS program, {\it effarea}, uses a point-source Ancillary Response File (ARF) and a surface brightness profile (SBP) to produce a MECS ARF for a concentric, on-axis region of radially symmetric, extended emission.  See \citet{fio} for details about this procedure.  We obtained an on-axis, point-source ARF from the Italian Space Agency (ASI) web-site and used the HRI data to create a SBP.  Using {\it effarea}, we created 12 ARFs for concentric regions with a width (inner to outer radius) of 0$\arcmin$.5 ($\sim\case{1}{2}$ of the  50\% power radius [PR$_{50}$]  of the MECS [1$\arcmin$.2 at 6.4 keV) \citep{fio}]. We assumed a linear response between channel number and energy bin and used the 12 concentric ARFs to create an exposure map for a circular region with a 6$\arcmin$ radius.  We wrote an algorithm based on the detector and sky pixels of the MECS2 source data to transform the MECS2 background events data from detector pixels to sky pixels.  This algorithm produced an error of no more 3 pixel sky coordinates for the MECS2 source data, corresponding to an error of $\leq$0$\arcmin$.4, which is much smaller than the MECS PR$_{50}$ of 1$\arcmin$.2 at 6.4 keV \citep{fio}.  We created a smoothed X-ray flux map (Figure \ref{fig1}) from the MECS2 events data (limited to the 1.5 - 10.5 keV range), using the corrected MECS2 background file (blank sky, taken at high galactic latitude) and ARF created exposure map.  The 6$\arcmin$ circular region contains the majority of cluster X-ray emission, while avoiding problems associated with the Beryllium window support structure \citep{boe}, and also limits the number of Faint Images of the Radio Sky at Twenty cm (FIRST) survey \citep{white} in our extraction region to seven. Figure \ref{fig1} also includes contours from the NRAO VLA Sky Survey NVSS data \citep{condon} which was smoothed with a Gaussian equal to its beam width of 45$\arcsec$, giving a resolution of $\sim$64$\arcsec$, and the Digital Sky Survey (DSS) image of the cluster.

We obtained {\it BeppoSAX} Low Energy Concentrator Spectrometer (LECS), MECS2, and MECS3 cleaned event files, from the ASI web-site, and extracted spectral files from these source event files. We used the 6$\arcmin$ on-axis spectral background files provided at the ASI web-site.  Using the procedure described above, we created a MECS ARF for a 6$\arcmin$ region of extended emission using {\it effarea}.  We used the on-axis Redistribution Matrix File (RMF), available for the MECS at the ASI web-site.

In the case of the LECS, we created an ARF and RMF by considering our extended region to be numerous point sources, and using a procedure similar to \citet{ett}.  The program, {\it lemat}, creates an ARF and a RMF for off-axis point-source observations of the LECS.  We used {\it lemat} to create ARFs and RMFs for $\sim$45 regions with 18 or more counts, and then created a count weighted ARF and RMF for the entire region.

Since the MECS2 and MECS3 extraction regions are identical, we fit them with all components tied.  However, since they have different responses, we fit them as separate files, rather than merging them.  Normalization cross calibration between the MECS2 and MECS3 is generally good ($\sim$3\%) according to the ASI web-site, but varies between the LECS and MECS depending on the position of the centroid \citep{fio}.  We, therefore, kept the normalizations between the LECS and MECS untied, even though the extraction regions were identical.

We used the HRI to determine the X-ray flux from the seven FIRST sources with in our 6$\arcmin$ extraction region.  Since we were looking for point sources in a region with extended X-ray emission, it was important to subtract both X-ray background and local, diffuse emission.  For our seven FIRST sources, we used 8$\arcsec$ regions to determine the source count rate.  In order to determine the proper background plus diffuse emission we extracted seven 8$\arcsec$ regions that were the same distance from the center of the diffuse emission as the source region.  The one exception to this was the cD galaxy, which is very close to the center of the emission($\sim$2$\arcsec$) (see Figure \ref{fig1}, FIRST source 1).  In order to determine the background and diffuse emission in the vicinity of this source we used six 8$\arcsec$ regions adjacent to the 8$\arcsec$ source region.  Unfortunately, there are no detections above the 3$\sigma$ level in the HRI for any of the FIRST sources, so that the HRI cannot provide constraints on their X-ray flux.  

We also used the HRI to identify possible X-ray sources that had not been detected in the current radio observations (i.e. radio quiet AGN).  Using algorithms to determine point sources in diffuse regions can be dangerous, because it is difficult to account for the diffuse emission.  Since we are looking for significant fluctuations in a field that is $\sim$10$^{6}$ square pixels, significance means $> \sim$1 in 10$^{6}$ which is $\sim$5$\sigma$.  Therefore, we argue any identifiable X-ray source should be visible above the X-ray background and diffuse emission.  In fact, we identified the same three sources identified by \citet{trin}, only one of which (identified by \citet{trin} as USNO-A1.0 source U1275\_09534024) is in our 6$\arcmin$ extraction region.  Since this source is close to the center of the diffuse emission ($\sim$0.9$\arcmin$), we used six regions adjacent to our extraction region to determine the X-ray background plus local diffuse emission behind this source.
 
\section{Results and Analysis}

\begin{figure}[tb]
\includegraphics[scale=0.4,angle=0]{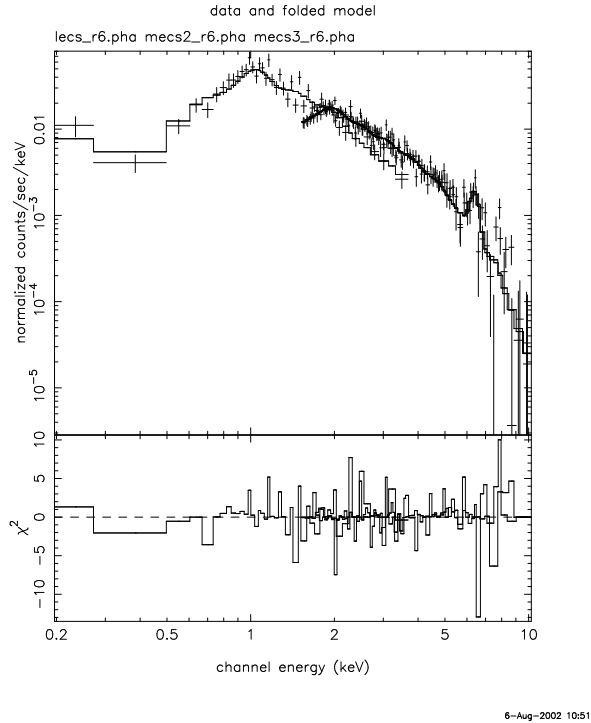}
\figcaption[f2.eps]
{Mekal fit to LECS, MECS2, \& MECS3 data.\label{fig3}}
\end{figure}

We fit the spectra with multiple component models (Table \ref{tbl-1}). Figure \ref{fig3} shows the plot of the single mekal \citep{mew01,mew02,kaa02,lied} fit.  Residuals are visible in the high end of the MECS data.  Although a two component mekal model  gives the best fit to the data, the high temperature component is unconstrained, and the two temperatures overlap within the 90\% confidence range, meaning that at the 90\% confidence level, this model is consistent with a single temperature.  In addition, by examining the temperature-luminosity relationship found by \citet{horn}, we find that the high temperature component produces a luminosity not consistent with its best fit temperature.  Our two component mekal model gives a luminosity of (1.1 - 31.9) $\times$ 10$^{42}$ {\it h$_{50}^{-2}$} ergs s$^{-1}$ for the best fit temperature of 2.88 keV, which is well below the luminosity $\sim$(1.6 - 2.0) $\times$ 10$^{44}$ {\it h$_{50}^{-2}$} ergs s$^{-1}$ predicted by \citet{horn} for this temperature.  Only the best fit temperature is considered, rather than the 90\% confidence range, because the 90\% upper limit is unconstrained and the 90\% confidence lower limit overlaps the low temperature component.

\begin{deluxetable}{lcccccc}[tb]
\tabletypesize{\scriptsize}
\tablewidth{0pt}
\tablecolumns{7}
\tablecaption{Spectral Fits of Regions. \label{tbl-1}}
\tablehead{\colhead{Model} & \colhead{kT$_{1}$ (keV)} & \colhead{Abundance} & \colhead{kT$_{2}$ (keV)} & \colhead{$\Gamma_{X}$} & \colhead{n$_{H}$ 10$^{20}$ cm$^{-2}$} & \colhead{$\chi^{2}$/dof}}
\startdata

mekal & 2.23$^{+0.09}_{-0.09}$ & 0.60$^{+.17}_{-0.15}$ & \nodata & \nodata & 4.1$^{+1.8}_{-1.2}$ & 239.6/188 \\
mekal+mekal & 2.88$^{+>97.12}_{-1.95}$ & 0.45$^{+0.18}_{-0.14}$ & 1.29$^{+0.65}_{-0.32}$ & \nodata & 4.7$^{+3.2}_{-1.5}$ & 230.9/185 \\
mekal+power law & 1.94$^{+0.38}_{-0.15}$ & 0.69$^{+0.32}_{-0.22}$ & \nodata & 1.7$^{+1.1}_{-1.2}$ & 4.4$^{+2.9}_{-1.6}$ & 231.5/185 \\
mekal+power law\tablenotemark{\dag} & 2.23$^{+0.12}_{-0.11}$ & 0.75$^{+0.39}_{-0.20}$ & \nodata & 2.8\tablenotemark{\dag} & 4.0$^{+2.4}_{-1.3}$ & 233.2/186 \\
\enddata
\tablenotetext{\dag}{Power law frozen at 2.8}
\end{deluxetable}

\begin{figure}[tb]
\includegraphics[scale=0.4,angle=0]{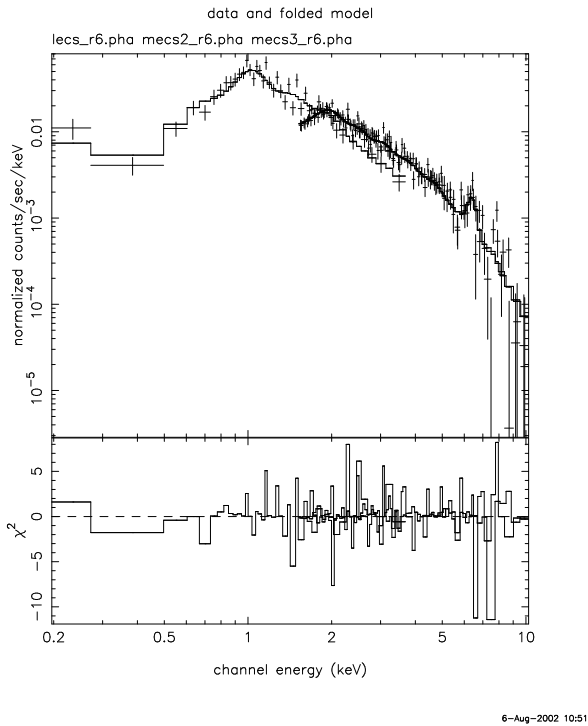}
\figcaption[f3.eps]
{Mekal+power-law fit to LECS, MECS2, \& MECS3 data.\label{fig4}}

\end{figure}

The addition of a power law component to the data gives a similar improvement to the fit (ftest probability = 90\% that power-law is needed)(Figure \ref{fig4}),  and gives a thermal luminosity of (2.2 - 51.6)$\times$ 10$^{42}$ {\it h$_{50}^{-2}$} ergs s$^{-1}$ for the best fit temperature of 1.95 keV, which is consistent with the results of \citet{horn} (($\sim$46 - 52) $\times$ 10$^{42}$ {\it h$_{50}^{-2}$} ergs s$^{-1}$).  The best fit temperature is used in comparing the two models (considering a temperature range will only extend the error bars so that the two luminosities will still overlap).  Our nonthermal detection in the MECS is at the 2.78$\sigma$ level, and gives $\Gamma_{X}$ = 0.4 - 2.8 with a flux of (3.4 - 25.7) $\times$ 10$^{-13}$ ergs cm$^{-2}$ s$^{-1}$ over 1.5 - 10.5 keV.  The large range of flux is because of the large uncertainty in both $\Gamma_{X}$ and the normalization.

In order to confirm that the nonthermal emission is diffuse, we used the HRI observation to try to determine the X-ray flux from point sources in our 6$\arcmin$ extraction region.  As discussed earlier, the seven FIRST sources were not detected in the HRI, so that we have no constraint on their X-ray emission.  U1275\_09534024, however, was detected at the $\sim$5$\sigma$ level in the HRI, but the lack of spectral information in the HRI data precludes the analysis of U1275\_09534024's spectrum.  We assumed a power law with $\Gamma_{X}$ equal to our best-fit value of 1.7, in order to see if it is a possible source of our nonthermal detection.  Without an identification the possibility exists that this source is variable in the X-ray. Since it is impossible to resolve it with the LECS or the MECS to determine its X-ray emission during the {\it BeppoSAX} observation, we can only assume that its X-ray flux does not vary above the 3$\sigma$ level.  We subtracted the X-ray background and diffuse emission from the HRI observation and determined that the flux from U1275\_09534024 is (0.1 - 10.2) $\times$ 10$^{-13}$ ergs cm$^{-2}$ s$^{-1}$ over 1.5 - 10.5 keV.  Although this result falls within the 90\% confidence range of the measured nonthermal emission, we point out that \citet{trin} noted that U1275\_09534024's X-ray-to-optical flux ratio is only marginally consistent with an AGN, and is more consistent with a star.  Therefore, it may not be producing any nonthermal emission at all.

\section{Diffuse Nonthermal Emission}

Using the {\it ROSAT} HRI observation of IC1262, and assuming our best fit $\Gamma_{X}$ of 1.7, we determine that U1275\_09534024 can account for the nonthermal emission we detected.  However, there is additional evidence that there is diffuse nonthermal emission.  By subtracting the point sources in NVSS and WENSS radio observations of our region, we find a radio excess of 154 mJy $\pm$7 at 1.4 GHz and 2301 mJy $\pm$41 at 0.33 GHz.

\begin{figure}[tb]
\includegraphics[scale=0.4,angle=0]{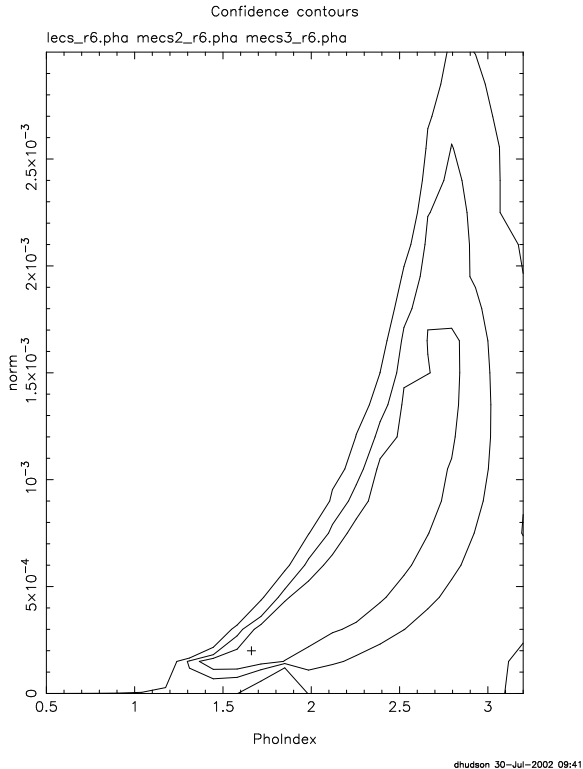}
\figcaption[f2.eps]
{Contour plot of $\Gamma_{X}$ versus Normalization.\label{fig2}}

\end{figure}

We note that the number of point sources detected in the Westerbork Northern Sky Survey (WENSS; 0.33 GHz) is different than the number in the NVSS(1.4 GHz),  however this maybe due to AGN self-absorption at low frequencies. Considering a simple power-law fit to the radio excess, we find a energy index, $\alpha_{r} \sim$1.8, which is consistent within 90\% confidence to the result found with the X-ray observation ($\Gamma_{X}$ = 2.8).  Examining the contour plot of $\Gamma_{X}$ versus normalization (Figure \ref{fig2}), we note that the normalization is much larger for steeper photon indices.  In fact, if the photon index is frozen at 2.8, the best fit model predicts a flux of (6.2 - 29.2) $\times$ 10$^{-13}$ ergs cm$^{-2}$ s$^{-1}$ over 1.5 - 10.5 keV. For this photon index, U1275\_09534024 cannot account for the flux we see.

Examining each of the nonthermal models more closely, we note that the temperature decreases for the addition of a free power law (see Table \ref{tbl-1}), and remains constant for the power law with a photon index constrained to the implied radio value ($\Gamma_{X}$ = 2.8).  For a power law with a free spectral index, the high energy photons are fitted with the power law, which causes a reduction in the temperature.  When the photon index is frozen at the radio implied value ($\Gamma_{X}$ = 2.8), the power-law provides a significant number of photons at low energies (in the MECS), preventing a decrease in the temperature.  By freezing the power law at $\Gamma$ = 2.8, and its normalization to the 90\% confidence lower-limit, the best fit temperature remains 2.23 keV.  In this case, the power-law fit simply provides photons for the high energy band of the spectrum.  If instead, the power-law normalization is frozen at its 90\% confidence upper-limit, the steep power law dominates at low energies, forcing the best-fit temperature up to 2.27 keV.  As with the other power-law models, the thermal model falls off at the high energy, where the power law begins to dominate again.  The 90\% confidence upper-limit for the normalization, with $\Gamma$ = 2.8 seems physically implausible for several reasons: the implied B field is $<$0.1 $\mu$G (see Section 5 below), the MECS spectrum is dominated by the power law and not the thermal component, the best-fit abundance is very high ($\sim$0.95), and a steep power law with a relatively high flux should also be visible in the LECS.  This implies that as with hot, rich clusters (such as Coma), a simple power law is probably not the best model for the nonthermal spectrum, but without more data, it is the only model we can construct.

\section{IC1262's Radio Halo}
The strongest evidence of diffuse, nonthermal, radio emission comes from the excess radio emission found after subtracting point sources from the NVSS and WENSS observations.  The 3$\sigma$ radio contour for the NVSS data in Figure \ref{fig1}, implies that the halo extends a distance of $\sim$6$\arcmin$ which corresponds to a distance of about 360 {\it h$_{50}^{-1}$} kpc (radius $\sim$180 {\it h$_{50}^{-1}$} kpc).  Although relativistic electrons from the cD galaxy (FIRST source 1 in Figure \ref{fig1}) could be leaking out, traveling at their Alfv\'{e}n velocity, they could only reach a distance of $<$8.0 {\it h$_{50}^{-1/4}$} kpc within their radiative lifetime ($\sim$10$^{8}$ years), implying that cosmic ray acceleration must occur in the intra-cluster medium.

\begin{figure}[tb]
\includegraphics[scale=0.4,angle=0]{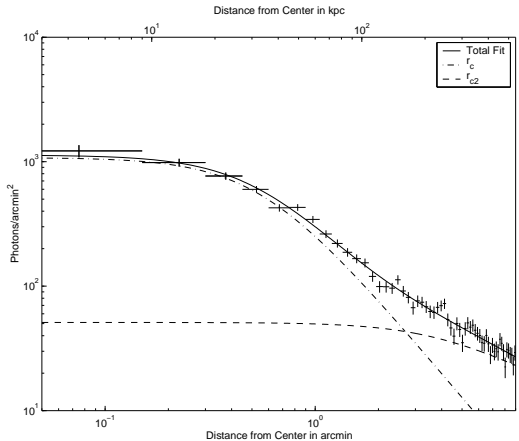}
\figcaption[f5.eps]
{Double King fit to HRI radial profile of IC1262 \label{fig_beta}}

\end{figure}

Radio halos are usually on the order of 0.8 - 1.2 {\it h$_{50}^{-1}$ }Mpc \citep{fer}.  The so-called mini-halos, which are similar in size to IC1262's radio-halo ($\sim$200 kpc \citep{gitti}), are observed in cooling flow clusters dominated by cD galaxies \citep{gitti}.  Since cooling flows can be seen as an increase in surface brightness over the diffuse emission in the center of a cluster, we fit the radial profile (from the HRI data) of IC1262 to a $\beta$ model.  Although a central peak was not explicitly visible in our fit to the surface brightness profile, a single $\beta$ model did not adequately fit the data.  Adding a second component to the $\beta$ model reduces $\chi^{2}$ by 30 for a reduction of two degrees of freedom.  This model (see Figure \ref{fig_beta}) gives an inner core radius of 31$\arcsec$ -  35$\arcsec$ (31 - 35 {\it h$^{-1}_{50}$} kpc) with a $\beta$ = 0.48 - 0.52, and an outer core radius of 213$\arcsec$ - 252$\arcsec$ (213 - 252 {\it h$^{-1}_{50}$} kpc) with a $\beta$ = 0.30 - 0.76.  Examining figure \ref{fig_beta}, we see that the central region dominates out to $\sim100$ {\it h$_{50}$} kpc, which is $\sim\case{1}{2}$ the size of the radio halo.  From this we conclude that the radio halo associated with IC1262 is not a mini-halo since it is much larger than the cooling flow.

In passing, we note that although the best fit $\beta$ for the outer core radius is 0.31, which is much flatter than expected for a cluster of galaxies, and is consistent with results found by \citet{hels} for loose groups of galaxies, within our 90\% confidence range, $\beta$ is consistent with the value of $\sim$0.66 associated with clusters of galaxies \citep{jones}.  We also point out that IC1262's diffuse gas has a temperature of $\sim$2 keV, which is hotter than for for any known group ($\sim$0.3 - 1.8 keV \citep{mul}), and that the outer core radius of $\sim$250 {\it h$_{50}$} kpc is typical for a cluster \citep{jones}.

Following the method outlined in \citet{henr}, and assuming a simple ICS power-law model ($\Gamma_{X}$ = 1.7), we determined the magnetic field for the 6$\arcmin$ extraction region.   Using the nonthermal, X-ray flux, radio excess, and best-fit photon index, we find an average cluster magnetic field of 0.01 - 0.87 $\mu$G for the NVSS, and 0.04 - 1.0 $\mu$G for the WENSS.  These magnetic fields are consistent with the typical range for an average cluster magnetic field (0.1 - 1.0 $\mu$G \citep{dol}), providing more evidence to the existence of diffuse emission.  The fact that the lower limits are $<$0.1$\mu$G implies that the nonthermal flux is probably less than the 90\% confidence upper limit.  \citet{petro}, however has demonstrated that using models other than simple power laws for the radio emission, higher magnetic fields are possible in clusters such as Coma, even if the measured nonthermal X-ray flux is considered to be created via the ICS process.

\section{Discussion}
IC1262 is a poor, cool cluster that shows some evidence of nonthermal X-ray emission. While, a similar claim can be made for another low mass system, the compact group HCG 62, IC1262 is distinguished in that it shows evidence for a radio halo, where HCG 62 does not \citep{fuka}. Currently, cluster mergers are the most popular model for the cosmic ray acceleration that leads to the production of diffuse nonthermal emission. IC1262 shows no direct X-ray evidence of a merger, such as elongated X-ray emission (as in the case of A754), or bi-modal X-ray morphology (as in the case of A1750).  There is, however, optical evidence that the cluster may have undergone a merger.  The cD galaxy has a red-shift of z = 0.0328 (9858 km s$^{-1}$)\citep{weg} compared to a red-shift of z = 0.0343 (10311 km s$^{-1}$) \citep{weg} for the cluster.  This gives a peculiar velocity, $\Delta$v, of 453 km s$^{-1}$, which is $>$3$\sigma$ above the dispersion of peculiar velocities (168$^{+41}_{-34}$ km s$^{-1}$) of cD galaxies found by  \citet{oeg}.  A line of sight merger could be the explanation for this unusually high peculiar velocity.  In addition, \citet{trin} analyzed the {\it ROSAT} HRI observation of IC1262 and found bright arc near the cD galaxy, which may indicate dynamic evolution of the cD galaxy due to a merger.

Our analysis has suggested that diffuse, nonthermal emission is associated with IC1262.  We find a 2.78$\sigma$ detection in the MECS, but no detection in the LECS within 90\% confidence.  This lack of detection in the LECS is probably because it is not as sensitive to hard photons as the MECS.  Although our 90\% confidence in $\Gamma_{X}$ and normalization indicate that U1275\_09534024 could be the reason for our nonthermal detection, the radio implies the existence of diffuse nonthermal emission.  When the diffuse radio excess is fit to a simple power law, it gives a spectral index, $\alpha_{r}$ $\sim$1.8.  When we use the radio to confine $\Gamma_{X}$ to 2.8, then the nonthermal, X-ray emission can not be accounted for with point sources.  We also reiterate that U1275\_09534024 may not even be emitting nonthermal, X-ray emission. A deeper radio observation will be able to confirm the existence of a radio halo and provide a spectral index that can be used to constrain $\Gamma_{X}$.  With a sensitivity up to $\sim$30 keV, the {\it Rossi X-Ray Timing Explorer} Proportional Counter Array(PCA) should be able to detect any diffuse, nonthermal, X-ray emission associated with IC1262, and also would be able to determine the variability, if any, of U1275\_09534024.

\acknowledgements
We would like to acknowledge and thank the National Science Foundation for its support and Dr. Tracy Clarke for her help with analysis of the radio data.


\end{document}